\documentclass[12pt]{article}

\usepackage{epsfig}
\usepackage{amssymb}
\usepackage{subfigure}
\usepackage{graphicx}
\usepackage{color}
\usepackage{cite}

\begin{document}

\baselineskip 6mm
\renewcommand{\thefootnote}{\fnsymbol{footnote}}


\newcommand{\nc}{\newcommand}
\newcommand{\rnc}{\renewcommand}



\newcommand{\tcb}{\textcolor{blue}}
\newcommand{\tcr}{\textcolor{red}}
\newcommand{\tcg}{\textcolor{green}}


\def\ba{\begin{array}}
\def\ea{\end{array}}
\def\be{\begin{eqnarray}}
\def\ee{\end{eqnarray}}
\def\nn{\nonumber\\}


\def\ct{\cite}
\def\la{\label}
\def\eq#1{(\ref{#1})}


\def\a{\alpha}
\def\b{\beta}
\def\g{\gamma}
\def\G{\Gamma}
\def\d{\delta}
\def\D{\Delta}
\def\e{\epsilon}
\def\et{\eta}
\def\ph{\phi}
\def\Ph{\Phi}
\def\ps{\psi}
\def\Ps{\Psi}
\def\k{\kappa}
\def\l{\lambda}
\def\L{\Lambda}
\def\m{\mu}
\def\n{\nu}
\def\th{\theta}
\def\Th{\Theta}
\def\r{\rho}
\def\s{\sigma}
\def\S{\Sigma}
\def\ta{\tau}
\def\o{\omega}
\def\O{\Omega}
\def\pr{\prime}


\def\half{\frac{1}{2}}
\def\goto{\rightarrow}

\def\na{\nabla}
\def\grad{\nabla}
\def\curl{\nabla\times}
\def\div{\nabla\cdot}
\def\pa{\partial}
\def\fr{\frac}

\def\bra{\left\langle}
\def\ket{\right\rangle}
\def\lb{\left[}
\def\lc{\left\{}
\def\ls{\left(}
\def\lp{\left.}
\def\rp{\right.}
\def\rb{\right]}
\def\rc{\right\}}
\def\rs{\right)}

\def\vac#1{\mid #1 \rangle}


\def\td#1{\tilde{#1}}
\def\check{ \maltese {\bf Check!}}


\def\Tr{{\rm Tr}\,}
\def\det{{\rm det}}
\def\text#1{{\rm #1}}


\def\bc#1{\nnindent {\bf $\bullet$ #1} \\ }
\def\ch {$<Check!>$ }
\def\ss {\vspace{1.5cm}}
\def\inf{\infty}

\begin{titlepage}

\hfill\parbox{5cm} { }

 
\vspace{25mm}

\begin{center}
{\Large \bf  Holographic Entanglement Entropy in Cutoff AdS}

\vskip 1. cm
   {Chanyong Park$^{a}$\footnote{e-mail : cyong21@gist.ac.kr}}

\vskip 0.5cm

{\it $^a$ Department of Physics and Photon Science, Gwangju Institute of Science and Technology,  Gwangju  61005, Korea}

\end{center}

\thispagestyle{empty}

\vskip2cm


\centerline{\bf ABSTRACT} \vskip 4mm

\vspace{0.5cm}

We investigate the holographic entanglement entropy of deformed conformal field theories which are dual to a cutoff AdS space. The holographic entanglement entropy evaluated on a three-dimensional Poincare AdS space with a finite cutoff can be reinterpreted as that of the dual field theory deformed by either a boost or $T \bar{T}$ deformation. For the boost case, we show that, although it trivially acts on the underlying theory, it nontrivially affects the entanglement entropy due to the length contraction. For a three-dimensional AdS, we show that the effect of the boost transformation can be reinterpreted as the rescaling of the energy scale, similar to the $T \bar{T}$ deformation. Under the boost and $T \bar{T}$ deformation, the $c$-function of the entanglement entropy exactly shows the features expected by the Zamoldchikov's $c$-theorem. The deformed theory is always stationary at a UV fixed point and monotonically flows to another CFT in the IR fixed point. We also show that the holographic entanglement entropy in a Poincare cutoff AdS space can reproduce the exact same result of the $T \bar{T}$ deformed theory on a two-dimensional sphere.

\vspace{2cm}

\end{titlepage}

\renewcommand{\thefootnote}{\arabic{footnote}}
\setcounter{footnote}{0}



\section{Introduction}

In the last decade, considerable attention has been paid to the entanglement entropy for figuring out a variety of quantum feature for strongly interacting systems. When describing a quantum phase transition appearing at zero temperature, the entanglement entropy may play a central role as an order parameter representing such a phase transition \cite{Calabrese:2009qy}. The entanglement entropy is also one of the good measures detecting the quantum correlation between quantum states. Despite a well-established definition, in general, it is a formidable task to calculate the entanglement entropy for interacting quantum field theories (QFT). Based on the AdS/CFT correspondence \cite{Maldacena:1997re,Gubser:1998bc,Witten:1998qj,Witten:1998zw}, it has been conjectured that the quantum entanglement entropy of a conformal field theory (CFT) can be understood by calculating the area of the minimal surface extended to the dual bulk geometry \cite{Ryu:2006bv,Ryu:2006ef,Hubeny:2007xt,Casini:2011kv,Lewkowycz:2013nqa,Nishioka:2009un,Blanco:2013joa,Park:2015afa}. This holographic method has been widely used to clarify a variety of quantum features even for strongly interacting systems.

Recently, it has been shown that a two-dimensional CFT deformed by an irrelevant $T \bar{T}$ deformation is solvable due to its integrable structure \cite{Zamolodchikov:2004ce,Smirnov:2016lqw}. This integrable quantum field theory can be taken into account as an effective field theory with a finite ultraviolet (UV) cutoff. Further, its dual gravity theory was investigated in Ref. \cite{McGough:2016lol}. The holographic dual of a $T \bar{T}$ deformation was identified with a one-dimensional higher AdS geometry cutting the asymptotic region off and placing a dual QFT on the AdS boundary located at a finite distance. This proposition has passed several quantitative checks by comparing the results of both QFT and its dual gravity \cite{Giveon:2017nie,Shyam:2017znq,Guica:2017lia,Giribet:2017imm,Kraus:2018xrn,Aharony:2018vux,Bzowski:2018pcy,Dubovsky:2018bmo,Donnelly:2018bef,Chakraborty:2018kpr,Babaro:2018cmq}. Moreover, the generalization of the $T \bar{T}$ deformation to a higher dimensional theory has been studied \cite{Bonelli:2018kik,Taylor:2018xcy,Hartman:2018tkw,Shyam:2018sro}.

As mentioned before, the dual of the $T \bar{T}$ deformed two-dimensional CFT is dual to the three-dimensional cutoff AdS space. From the holography point of view, cutting the asymptotic region off is regarded as shifting the AdS boundary from infinity to a finite distance with an appropriate Dirichlet boundary condition. This radial evolution of the dual gravity can be described by the Hamilton-Jacobi equation in which the Hamilton constraint is associated with the renormalization group (RG) flow equation of the dual QFT \cite{deBoer:1999tgo,Balasubramanian:1999re,deHaro:2000vlm,Skenderis:2002wp,Lee:2017nma}. Intriguingly, it has been shown that the RG flow equation of the $T \bar{T}$ deformed theory is perfectly matched to the Hamilton-Jacobi equation of the dual gravity with a finite cutoff  \cite{McGough:2016lol,Donnelly:2018bef}.

In this work, we investigate how the boost and scale transformation affect the entanglement entropy of a CFT. Intriguingly, we found that the entanglement entropy of a  boosted system for a two-dimensional QFT can lead to the similar effect to the $T \bar{T}$ deformation and that the dual gravity is again described by a three-dimensional cutoff AdS space.  At first glance, the boost and scale transformations look trivial because they are just elements of the conformal symmetry group. However, it is not true for the entanglement entropy, even though the underlying theory is conformal. The reason is as follows. In order to define the entanglement entropy, we first divide a total system into two subsystems by taking an appropriate entangling surface. Then, the quantum entanglement between states of two subsystems is described by the entanglement entropy across the entangling surface. In this case, although the boost and scale transformation act trivially on the underlying theory, their action on the entanglement entropy becomes nontrivial because the boost and scale transformation are not well-defined local transformations in the subsystem bounded by the fixed entangling surface. In addition, the boost transformation generally leads to the length contraction which can change the size and shape of the entangling surface. Due to these reasons, it would be interesting to study how the entanglement entropy is affected by the boost transformation.
 
Intriguingly, we have found that the boost transformation in the boundary direction in the three-dimensional AdS space can be reinterpreted by rescaling the radial coordinate. Remembering that the dual two-dimensional CFT is defined at the AdS boundary, rescaling the radial coordinate is equivalent to moving the AdS boundary located at an infinity to a finite distance. This is very similar to the $T \bar{T}$ deformation described above. Applying the holographic entanglement entropy formula, we found that the $c$-function and its RG flow of the boosted entanglement entropy show very similar feature to those of the $T \bar{T}$ deformation. We further showed that the $c$-function of the boosted entanglement entropy exactly satisfies the Zamolodichikov's $c$-theorem \cite{Zamolodchikov:1986gt}. The $c$-function of the boosted entanglement entropy is stationary at a UV fixed point and monotonically decreases along the RG flow. At an IR fixed point, finally, we found that it arrives at another CFT, although we do not clearly understand what the IR CFT is. We further investigated how the rescale of the energy caused by the boost transformation affects the mutual information of two subsystems separated by a certain distance. We found that the quantum correlation length between two subsystems becomes shorter as the energy scale observing the dual QFT is getting lower. 

We also investigated the boost transformation of higher dimensional theories. Our study showed that reinterpretation of the boost transformation as the energy rescaling seems to be possible only for a three-dimensional AdS space and its two-dimensional CFT. Despite this result, the boost transformation still nontrivially acts on the entanglement entropy even in higher dimensional theories. The boosted entanglement entropy for higher dimensional theories leads to the change of the size and shape of the entangling region. Through the explicit calculation, we showed that the boost transformation modifies the disk-shaped entangling region to the ellipsoidal one and that the modified leading contribution to the resulting entanglement entropy still satisfies the area law with a modified UV divergent term. Lastly, we showed that the entanglement entropy of the $T \bar{T}$ deformed CFT on a two-dimensional sphere is perfectly matched to the holographic entanglement entropy calculated in a three-dimensional Poincare AdS space with a finite cutoff. We also showed that the RG flow of this $T \bar{T}$ deformed CFT gives rise to the almost same feature as the that of the boost transformation up to a numerical factor.
 
The rest of this paper is as follows: In Sec. 2, we study the entanglement entropy of a two-dimensional boosted system and show that the boost can be reinterpreted as the shift of the UV cutoff. This fact allows us to understand the RG flow of $c$-function from the UV fixed point to another IR fixed point. In Sec. 3, we consider the effect of the boost in a three-dimensional CFT field theory. In this case, the boost changes a ball-shaped entangling region into an ellipsoidal one due to the length contraction. In Sec. 4, we take into account a cutoff AdS$_3$ in a Poincare patch and reproduce the exact same entanglement entropy of the $T \bar{T}$-deformed theory on ${\bf S}^2$. Finally, we finish this work with some concluding remarks in Sec. 5.


\section{ Entanglement entropy of a boosted CFT}

Following the AdS/CFT correspondence \cite{Maldacena:1997re,Gubser:1998bc,Witten:1998qj,Witten:1998zw}, the dual gravity of a two-dimensional CFT is given by a three-dimensional AdS space with the following metric 
\be			\la{met:unboosted}
ds^2 = \fr{R^2}{z^2} \ls - dt^2 +  d x^2   + dz^2 \rs ,
\ee
where $R$ denotes a radius of the AdS space and the boundary is located at $z=0$. At the boundary, the AdS metric reduces to a two-dimensional Minkowski space represented by ${\bf R} ^{1,1}$. In this case, the isometry group of a three-dimensional AdS space is given by $SO(2,2)$, while the isometry group of the boundary space  becomes a two-dimensional Poincare group, $ISO(1,1)$. If a theory living on the boundary is conformal, the symmetry group of the boundary theory is enhanced to $SO(2,2)$. This conformal group has a one-to-one map to the isometry group of the AdS space. The matching of two symmetry groups provides one of the evidence for the holography. In particular, the dilatation of the conformal symmetry is realized as the invariance of the bulk metric under the following scale transformation 
\be
z \to \l z \ , \quad t \to \l t  \ , \quad   {\rm and} \quad  z \to \l z  .
\ee
Since the conformal field theory includes a Lorentz symmetry, the boundary theory is also invariant under a boost transformation. The same thing is also true on the dual gravity side. Parameterizing a boost transformation in the $x$-direction as
\be  	\la{rel:boost}
\left(
\begin{array}{c}
 dt'     \\
dx'   
\end{array}
\right) =
\left(
\begin{array}{cc}
\cosh \b &  - \sinh \b   \\
- \sinh \b & \cosh \b 
\end{array}
\right)
\left(
\begin{array}{c}
 dt    \\
 dx     
\end{array}
\right) ,
\ee
it becomes manifest that the AdS metric is invariant under this boost transformation.

These scale and boost transformations of a CFT give rise to only a trivial result because they are elements of the symmetry group. In this case, we must notice that the CFT is defined in the entire region of ${\bf R} ^{1,1}$, where the symmetry group is closed. However, it is not the case for the entanglement entropy. In order to define the entanglement entropy, we first need to divide the entire space into two subspaces or subsystems. The border of them is called an entangling surface. If the size of the subsystem bounded by the entangling surface is finite, the scale and boost transformations do not provide well-defined local transformations of the subsystem. This is because any event of the subsystem can go to the outside of the subsystem by the scale and boost transformations. Consequently, the scale and boost transformations nontrivially act on the entanglement entropy. In other words, they can modify the size or shape of the subsystem and lead to an additional and nontrivial contribution to the entanglement entropy.

When we take a specific entangling surface to evaluate the entanglement entropy, the existence of a fixed entangling surface breaks the boost symmetry, as mentioned before. In the holographic setup, the entanglement entropy is determined by the area of the minimal surface which lies in a constant-time hypersurface \cite{Ryu:2006bv,Ryu:2006ef}. For more details, we divide the boundary space into two parts, a subsystem and its complement. In this case, the subsystem must be chosen by a space-like region. This implies that we can set $dt=0$ in \eq{met:unboosted} which determines the constant-time hypersurface of the dual AdS geometry. Then, the minimal surface lies on the constant-time hypersurface and anchored to the entangling surface defined at the boundary. 

In the holographic setup, the configuration of the minimal surface is determined by the radial coordinate given as a function of the boundary spatial coordinate, $z(x)$. If we take a subsystem in the range of
\be			\la{range:subsystem}
-l/2 \le x \le l/2,
\ee
the holographic entanglement entropy is governed by
\be
S_E = \fr{R}{4 G} \int_{-l/2}^{l/2} dx \ \fr{\sqrt{1 + z'^2}}{z}  ,
\ee
where the prime means a derivative with respect to $x$. Solving the equation of motion derived from the action allows the following exact solution
\be			\la{solution:minimalsurface}
z(x) = \sqrt{ \fr{l^2}{4}  - x^2} .
\ee
After plugging this solution back into the action, performing the integral gives rise to the well-known entanglement entropy of a two-dimensional CFT \cite{Ryu:2006bv}
\be
S_E =  \fr{c}{3} \log \fr{l}{\e} ,
\ee 
where the central charge of the CFT is related to the Newton constant of the dual gravity, $c=3 R/(2 G)$  \cite{Brown:1986nw}, and $\e$ is introduced as a UV cutoff to regularize a UV divergence.

Now, let us move to a boosted system. In order to distinguish the boosted system from the unboosted one, we use a primed coordinate $x'$ for a boosted system and an unprimed coordinate $x$ for an unboosted system, respectively. If the system is boosted with a velocity $v$, the boost in \eq{rel:boost} is characterized by
\be
\cosh \b= \frac{1}{\sqrt{1 - v^2}}  ,
\ee 
where the natural unit, $c=1$, was used. In the CFT defined on the entire boundary space, as mentioned before, the action of the boost is trivial so that the resulting metric of the boosted system is again given by an AdS metric described by the boosted coordinate
\be
ds^2 = \fr{R^2}{z^2} \ls - dt'^2 +  d x'^2   + dz^2 \rs .
\ee
Here the radial coordinate $z$ is unchanged because it is orthogonal to the boost transformation. Now, we take a constant-time hypersurface ($dt'=0$) to calculate the holographic entanglement entropy of the boosted system. In this case, the simultaneity of the boosted system gives rise to a relation between the time and space of the unboosted system
\be
dt = \tanh \b \ dx .
\ee
In addition, the boosted coordinate $x'$ defined on the constant-time hypersurface is related to the unboosted one $x$ through the following relation
\be
dx' = \fr{dx}{\cosh \b}   .
\ee
This is nothing but the length contraction appearing in the boosted system. If the subsystem size is defined as \eq{range:subsystem} in the unboosted system, the range of it in the boosted system is parameterized as
\be
-l'/2 \le x \le l'/2,
\ee
where $l' = l /\cosh \b =  l \ \sqrt{1- v^2} $ is always shorter than $l$ due to the length contraction. Using this fact, the entanglement entropy of the boosted system is given by
\be
S'_E = \fr{R}{4G} \int_{-l'/2}^{l'/2} dx' \ \fr{1}{z} \sqrt{1 + \ls \fr{\pa z}{\pa {x'}} \rs^2} .
\ee
This formula shows that the entanglement entropy of the boosted system has the same form as the unboosted one only except the size change of the subsystem caused by the length contraction. The appearance of the same entanglement entropy form is due to the invariance of the underlying theory under the boost transformation. As a consequence, the resulting entanglement entropy becomes
\be  	 \la{result:boosted0}
S'_E &=&  \fr{c}{3} \log \fr{l'}{\e}   \nn
         & =&    \fr{c}{3} \log \fr{l}{\e} +  \fr{c}{6} \log (1- v^2) .
\ee 
Here the first term is the entanglement entropy of the unboosted CFT, whereas the second term represents the contribution from the boost. When $v \to 0$, as expected, \eq{result:boosted0} reproduces the result of the unboosted CFT. Furthermore, this result shows that the boost transformation nontrivially acts on the entanglement entropy with providing an additional contribution.

Intriguingly, the additional correction caused by the boost can be reinterpreted as the shift of the UV cutoff similar to the $T \bar{T}$ deformation. To clarify this point, let us first consider a constant-time hypersurface of the boosted AdS metric. For $dt'=0$, the induced metric on the constant-time hypersurface is given by
\be			\la{indmetric:boosted}
ds_{in}^2 =\fr{R^2}{z^2} \ls dx'^2 +  dz^2 \rs  = \fr{R^2}{z^2} \ls \fr{dx^2 }{\cosh^2 \b}  + d z^2 \rs  .
\ee
In the unboosted system,  it is possible to reinterpret the boost as the rescale of the radial coordinate. More precisely, after introducing a new radial coordinate $\bar{z}$
\be			\la{coord:new}
\bar{z} = z \cosh \b , 
\ee
rewriting the induced metric leads to
\be			\la{metric:rescale}
ds_{in}^2 =  \fr{R^2}{\bar{z}^2} \ls  dx^2   + d \bar{z}^2 \rs ,
\ee
which is again an AdS space. Noting that the radial coordinate of the AdS space is matched to the energy scale of the dual field theory, the metric in \eq{metric:rescale} indicates that the boost transformation is associated with rescaling the energy of the dual field theory. Therefore, the  boosted dual field theory is defined at the lower energy scale than the unboosted one. This feature becomes manifest when we calculate the entanglement entropy.

With the induced metric in \eq{metric:rescale}, the entanglement entropy formula is given by
\be
\bar{S}_E = \fr{R}{4G} \int_{-l/2}^{l/2} dx \ \fr{R}{\bar{z}} \sqrt{1 + \ls \fr{\pa \bar{z}}{\pa {x}} \rs^2} .
\ee
The configuration of the minimal surface in this system is described by
\be					\la{solution:rescaled}
\bar{z}  = \sqrt{ \fr{l^2}{4}  - x^2}  .
\ee
If we introduce $\bar{\e}$ as a UV cutoff of the $\bar{z}$-coordinate, the resulting entanglement entropy in the $(x,\bar{z})$-system yields
\be		\la{result:rescale0}
\bar{S}_E =  \fr{c}{3} \log \fr{l}{\bar{\e}} .
\ee
From \eq{coord:new}, $\bar{\e}$ is related to the UV cutoff of the $z$-coordinate, $\e$. This relation can be understood as the rescale of the UV cutoff, $\bar{\e} = \e \cosh \b$. Using this relation, the entanglement entropy \eq{result:rescale0} derived in the $(x, \bar{z})$-system is exactly the same as \eq{result:boosted0} derived in the $(x', z)$-system. In order to interpret the result in \eq{result:rescale0} correctly, it must be noted that $l$ is the subsystem size measured at the energy scale of $\e$. At the energy scale of $\bar{\e}$, the size of the subsystem $\bar{l}$ must satisfies 
\be			\la{relation:oldnewsubsystme}
l = \bar{l} \sqrt{1+  \fr{4 \bar{\e}^2}{\bar{l}^2} }  .
\ee
This relation naturally appears because the configuration of the minimal surface is given by a semicircle. In the high energy limit ($\bar{\e}/\bar{l} \ll 1$), the entanglement entropy measured at the energy scale $\sim 1/\bar{\e}$ has the following expansion
\be		\la{result:boosted}
\bar{S}_E  = \fr{c}{3} \log \fr{\bar{l} }{\bar{\e}}  +  \fr{ 2  c \bar{\e}^2}{ 3  \bar{l}^2 }   + {\cal O} \ls \bar{\e}^4 \rs .
\ee
This result, as will be seen, has a similar form to the entanglement entropy modified by a $T \bar{T}$ deformation. Intriguingly, our result shows that the boost transformation as well as the $T \bar{T}$ deformation of a two-dimensional CFT can be reinterpreted as the shift of the UV cutoff to the IR direction. From the dual gravity viewpoint, the shift of the UV cutoff is dual to a cutoff AdS$_3$ space whose boundary is located at $\bar{\e} =\e \cosh \b$. Similar to the  $T \bar{T}$ deformation, the boost transformation of the entanglement entropy gets rid of the energy region higher than $1/\bar{\e}=\sqrt{1-v^2}/\e$. Especially, when  the boosting velocity increases, the UV cutoff moves to the deeper IR region. On the other hand, the CFT results are reproduced when $v \to 0$.

Since $1/\bar{\e}$ plays a role of the energy scale observing the dual field theory, we can think of the RG flow of a $c$-function which represents an effective degrees of freedom of the dual field theory. For the undeformed CFT, the central charge is defined as a derivative of the entanglement entropy with respect to the subsystem size \cite{Liu:2012eea,Myers:2012ed,Casini:2012ei,Liu:2013una,Park:2018ebm,Narayanan:2018ilr}
\be
c = 3 l \fr{\pa  S_E}{\pa l} .
\ee
Generalizing this formula to the deformed CFT, the $c$-function of the deformed theory may be written as 
\be			\la{definition:cfunction}
\bar{c} (\bar{\e})=3 \bar{l}  \fr{\pa  \bar{S}_E}{\pa \bar{l}} = c \ls 1  -   \fr{ 4\bar{\e}^2}{\bar{l}^2 } + \cdots \rs .
\ee
This $c$-function approaches the central charge of the undeformed CFT when $\bar{\e} \to 0$, as expected. In addition, the nontrivial $\bar{\e}$ dependence represents deviation from the CFT during the RG flow process. From this $c$-function derived from the entanglement entropy, for $\bar{l} \gg \bar{\e}$, the RG flow of the $c$-function is described by
\be			\la{result:RGflow}
\fr{\pa \bar{c}}{\pa \log \bar{\e}} \approx  - \fr{ 8\bar{\e}^2}{\bar{l}^2 }    < 0.
\ee
In this case, the increase of $\bar{\e}$ represents the direction of the RG flow. The negative value of \eq{result:RGflow} indicates that the $c$-function  monotonically decreases along the RG flow, which is consistent with the Zamolodchikov's $c$-theorem \cite{Casini:2004bw,Casini:2006es,Myers:2010xs,Myers:2010tj}.

Now, let us investigate further the $c$-function in a deep IR region. The reinterpretation of the boost transformation as the rescale of the energy yields the following exact entanglement entropy, which is valid in the entire energy scale,
\be		
\bar{S}_E =  \fr{c}{3} \log \fr{\bar{l}}{\bar{\e}}   +   \fr{c}{3} \log  \sqrt{1+  \fr{4 \bar{\e}^2}{\bar{l}^2} }   .
\ee
Note that in this relation the UV cutoff moves to the deep IR region as $v$ increases. Applying the formulas in \eq{definition:cfunction} and \eq{result:RGflow}, we finally obtain the $c$-function and its RG flow caused by the boost transformation
\be
\bar{c} (\bar{\e}) &=& \fr{c \ \bar{l}^2}{\bar{l}^2 + 4  \bar{\e}^2 } , \la{result:cfunBD} \\
\fr{\pa \bar{c}}{\pa \log \bar{\e}}  &= & -  \fr{ 8 c \ \bar{l}^2 \ \bar{\e}^2}{ \ls \bar{l}^2 + 4  \bar{\e}^2 \rs^2 }  < 0.  \la{result:RGflowBD}
\ee
Since $c$ is always positive, the negativity in \eq{result:RGflowBD} indicates that the RG flow of the $c$-function monotonically decreases in the entire range of the RG scale. Moreover, \eq{result:cfunBD} shows that the $c$-function has the central charge of the undeformed CFT in the UV limit ($\bar{\e} \to 0$) and monotonically decreases to be zero in the IR limit ($\bar{\e} \to \infty$). It is worth noting that the $c$-function of the deformed theory is stationary at the UV fixed point. Furthermore, the obtained result shows that there exists an IR fixed point where the RG flow becomes zero and the $c$-function reduces to a constant value, $\bar{c}=0$. This is the exactly expected feature of the RG flow (see Fig. 1), although it is not clear what kind of CFT appears in an IR fixed point.

\begin{figure}
\begin{center}
\vspace{0cm}
\hspace{-0.5cm}
\subfigure{\label{fig1a} \includegraphics[angle=0,width=0.5\textwidth]{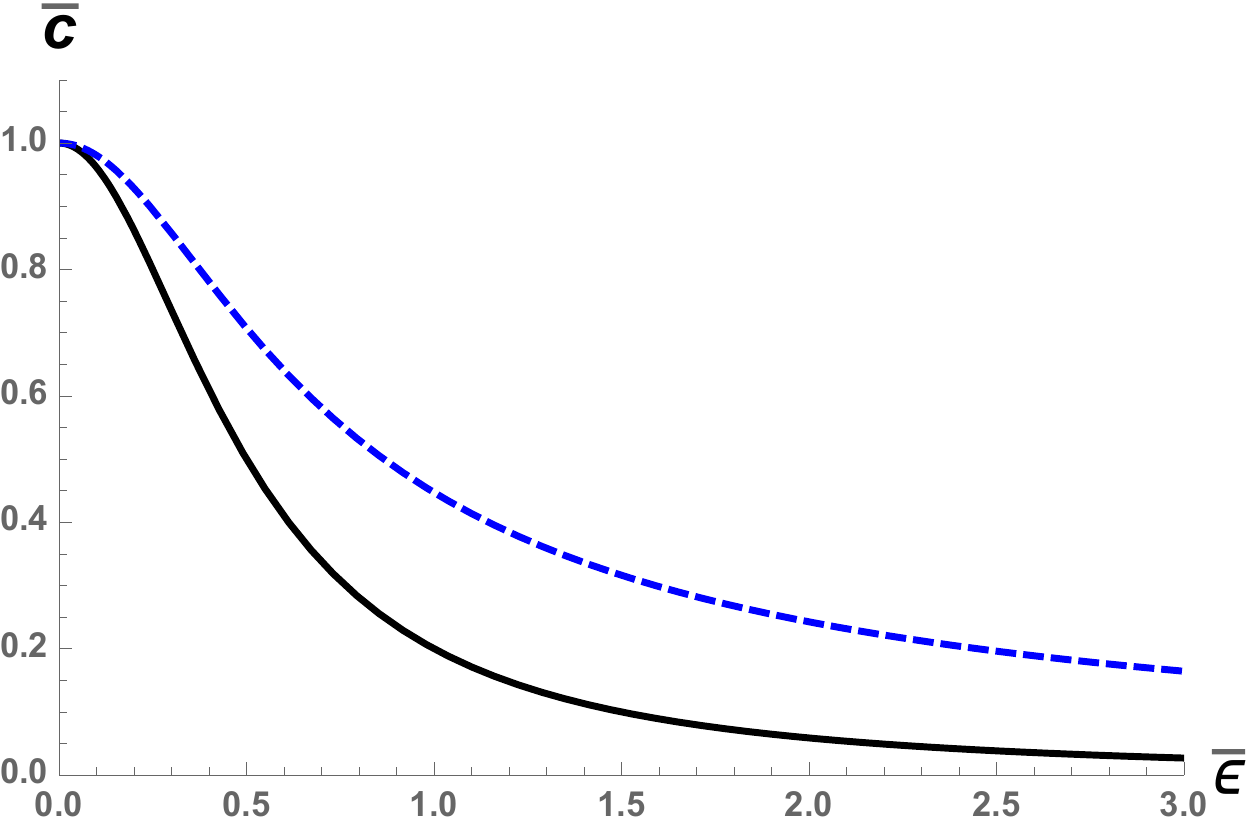}}

\caption{\small  The energy scale dependence of the $c$-function when a two-dimensional CFT is deformed by either a $T \bar{T}$ deformation (blue-dashed) or a boost transformation (black-solid).}
\label{fig1}
\end{center}
\end{figure}

We further take into account a mutual information and its energy scale dependence. When two subsystems having the same subsystem size $\bar{l}$ are far from each other with distance $h$, the mutual information is given by \cite{Casini:2006ws,Swingle:2010jz,Cardy:2013nua,Fischler:2012uv}
\be
I &=& 2 \bar{S}_E(\bar{l}) - \bar{S}_E(2 \bar{l} + h) - \bar{S}_E(h) \nn
&=& \fr{c}{6}  \log \lb \frac{\left(\bar{l}^2+4 \bar{\epsilon}^2\right)^2}{\left(h^2+4 \bar{\epsilon} ^2\right) \ls  (2 \bar{l} + h)^2 + 4 \bar{\epsilon}^2 \rs}\rb .
\ee
In the UV limit ($\bar{\e} \to 0$), there is no mutual information when the distance of two subsystems becomes larger than a critical distance, $h_c = ( \sqrt{2} - 1) \bar{l}$. If the underlying theory is conformal without any deformation, this is  true in the entire energy scale. However, if a CFT is deformed, the deformation can provide a nontrivially effect relying on the RG scale. Since the boost transformation can be regarded as the nontrivial rescale of the energy, we can easily expect that the mutual information is also affected by the boost. To understand how the mutual information of the deformed CFT changes, we calculate the change of the mutual information with respect to $h$ with rescaling the energy scale
\be
\fr{d I}{d h} = -\frac{2 c \left(\bar{l}+h\right) \left(2 h \bar{l}+4 \bar{\epsilon }^2+h^2\right)}{3 \left(4 \bar{\e}^2+h^2\right) \left(\left(2 \bar{l}+h\right)^2+4 \bar{\epsilon }^2\right)} < 0 .
\ee
This result, as expected, indicates that the mutual information monotonically decreases as the distance $h$ is getting farther. In addition, the boost  makes the critical distance shorter 
\be
h_c =  \ls \sqrt{2 \ls 1 - \fr{4 \bar{\e}^2 }{\bar{l}^2} \rs } - 1 \rs  \bar{l}   .
\ee
Consequently, the result shows that, if $h > h_c$, there is no mutual information between two subsystems, and that the critical distance becomes shorter along the RG flow.

\section{Boost in a higher dimensional AdS space}

Until now, we have studied the entanglement entropy of a two-dimensional CFT deformed by the boost transformation. In this case, we showed that the boost can be reinterpreted as the shift of the UV cutoff.  On the dual gravity side, such a rescale of the UV cutoff corresponds to the shift of an AdS boundary with resulting in a cutoff AdS$_3$. Now, we can ask whether the reinterpretation of the boost as the shift of the UV cutoff or the rescaling of the energy is a general feature applicable to higher dimensional CFTs. Unfortunately, the answer is negative, as will be seen later. As a consequence, the reinterpretation of the boost transformation is not a general but specific feature appearing only in a two-dimensional CFT and its dual gravity. Even in this case, it would be important and interesting to investigate how the boost modifies the entanglement entropy of a higher dimensional theory. For example, if we consider a ball-shaped entangling region as a subsystem, the boost can modify the ball-shaped region into an ellipsoidal one due to the length contraction in the boosted direction. As shown in this example, the boost transformation can provide more opportunities to figure out the entanglement entropy defined on a complicated shape of the entangling region.

The entanglement entropy calculation discussed in the previous section can be easily generalized to higher dimensional cases. For simplicity,  let us focus on a three-dimensional CFT which is dual to a four-dimensional AdS space
\be
ds^2 = \fr{R^2}{z^2} \ls - dt^2 +  d x_1^2 + dx_2^2   + dz^2 \rs .
\ee
Let us take into account a boost in the $x_1$-direction. Following the same strategy used in the previous section, the induced metric on a constant-time  hypersurfce reduces to
\be			
ds_{b}^2 = \fr{R^2}{z^2}   \ls {dx'_1}^2 + d x_2^2 + d z^2 
 \rs= \fr{R^2}{z^2} \ls  \fr{dx_1^2}{\cosh^2 \b}   + d x_2^2  + d z^2 \rs .
\ee
This result shows that unlike the previous AdS$_3$ case, the boost transformation cannot be generally reinterpreted as the rescaling of the radial coordinate $z$. This fact indicates that the reinterpretation of the boost transformation as the shift of the UV cutoff is a specific feature of a two-dimensional theory and its dual gravity.

Now, we consider a disk-shaped subsystem in the unboosted system, whose boundary is parameterized by a circle
\be
x_1^2 + x_2^2 = l^2 .
\ee
In the boosted system, the length contraction in the $x_1$ direction squeezes the circle and the shape of the entangling surface is modified into an ellipse satisfying
\be			\la{equation:ellipse}
\cosh^2 \b \ {x'_1}^2 + x_2^2 = l^2  .
\ee
This shows that the entanglement entropy of the boosted system is the same as the entanglement entropy of the unboosted one whose entangling surface is given by an ellipse shown in \eq{equation:ellipse}. In order to calculate the entanglement entropy defined in the elliptic region, we introduce a polar coordinate, $\r = \sqrt{{x'_1}^2 + x_2^2}$. Then, the metric on the constant-time hypersurface becomes
\be
ds_{b}^2 = \fr{R^2}{z^2}   \ls {d \r}^2 + \r^2 d \th^2 + d z^2 \rs .
\ee
If $z$ does not depend on $\th$, the action governing the holographic entanglement entropy is given by
\be
S_E = \fr{1}{G} \int_0^{\pi/2} d \th \int_0^{\bar{\r}-\d} d \r \ \fr{\r}{z^2} \sqrt{1 + \ls \fr{\pa z}{\pa \r} \rs^2}  ,
\ee
where $\int_0^{2 \pi} d \th = 4 \int_0^{\pi/2} d \th$ was used due to the symmetry of the elliptic region. Above $\bar{\r}$ in the $\r$ integration indicates a maximum value of $\r$ and $\d$ ($\d \ll 1$) was introduced to regularize the UV divergence. Since the upper bound of the $\r$-integral must be located on the ellipse, $\bar{\r}$ is given by a function of $\th$
\be
\bar{\r} = \fr{l}{\sqrt{\cosh^2 \b \ \cos^2 \th + \sin^2 \th}} 
\ee
This relation indicates that the semi-major axes $\bar{\r} = l$ appears at $\th=\pi/2$, whereas the semi-minor axes $\bar{\r} = l/\cosh \b$ is located at $\th=0$. Then, the solution of the equation of motion can be written as the following form 
\be			\la{solution:msurface}
z (\r)= \sqrt{\bar{\r}^2 - \r^2} .
\ee
Here it is natural to impose the boundary condition with $z(\bar{\r}) = 0$ because the entangling surface must live on the boundary space.

Plugging the solution into the action and performing the integral over $\r$ reduce the entanglement entropy to
\be 
S'_E = \fr{1}{G} \int_0^{\pi/2} d \th  \ls  \fr{\bar{\r}}{\sqrt{\bar{\r}^2 - (\bar{\r} - \d)^2}}- 1\rs .
\ee
From the solution in \eq{solution:msurface}, the UV cutoff $\d$ is related to the UV cutoff in the $z$-direction, $\e$,
\be
\d = \fr{\e^2}{2 l}  +{\cal O} \ls \e^4 \rs.
\ee
As a result, the resulting entanglement entropy gives rise to the following analytic expansion in the limit of $v \ll 1$
\be
S'_E = \fr{\pi (8 - v^2)}{16 G} \fr{l}{\e} - \fr{\pi}{2 G}  + {\cal O} \ls \e \rs .
\ee
Here, the fact that the leading term of the entanglement entropy is proportional to $l$ indicates the area law of the entanglement entropy, as expected. More precisely, the entangling surface of the boosted system is not a circle but ellipse so that $l$ in the above formula must be represented as the semi-major axes of the ellipse instead of the radius of the circle. For the ellipse described by \eq{equation:ellipse}, the eccentricity is the same as the boost velocity and the circumference of the ellipse is given by an elliptic integral of the second kind, ${\cal E}(v)$. 
\be
l_e \equiv 4 l {\cal E}(v) = 4 l \int_0^{\pi/2} d \th \ \sqrt{1 - v^2 \sin^2 \th} .
\ee
Rewriting the above entanglement entropy in terms of the variables of the ellipse we finally obtain
\be
S'_E = \fr{\pi (8 - v^2)}{16 G} \fr{l_e}{4 {\cal E}(v)  \e} - \fr{\pi}{2 G}  - \frac{\pi  \left(8-v^2\right) }{128 G } \fr{4 {\cal E}(v)  \e}{l_e}+ {\cal O} \ls \e^3 \rs .
\ee
Again, in the limit of $v \ll1$, the entanglement entropy becomes
\be			\la{result:freeenergy}
S'_E = \ls 1 + \fr{v^2}{8} + {\cal O} (v^4) \rs \fr{l_e}{4 G \e}  - \fr{\pi}{2 G}  - \frac{\pi^2 \e  }{8 G l_e }.
\ee
This result shows that the UV divergence is modified because the entanglement surface changes due to the boost transformation. Despite the change of the entangling surface, the result indicates that the leading contribution to the entanglement entropy still satisfies the area law. It was known that the constant term in the above entanglement entropy is independent of the regularization and corresponds to the free energy of the dual CFT defined on the Euclidean $S^3$ space \cite{Casini:2012ei,Pufu:2016zxm}. The result indicates that the boost transformation we took does not modify the constant free energy in the UV limit ($\e \to 0$).

\section{Entanglement entropy deformed by a $T \bar{T}$ operator}

Now, let us discuss a $T \bar{T}$ deformation.  In a two-dimensional CFT, a $T \bar{T}$ deformation corresponds to an irrelevant operator and a deformed CFT can be described by \cite{McGough:2016lol}
\be
{\cal S}_{QFT} = {\cal S}_{CFT} - \m \int d^2 x \sqrt{-g} \  T \bar{T} ,
\ee
where $T=T_{++}$ and $\bar{T}=T_{--}$ are defined with a null coordinate, $x^{\pm}= x^0 \pm x^1$. According to the dimension counting, $\m$ is a dimensionful coupling with a mass dimension $-2$. In general, the value of a dimensionful coupling crucially relies on the energy scale observing the theory. Because of the absence of other dimensionful parameter, the value of $\m$ must have a direct relation to the energy scale like  $\m \sim \bar{\e}^2$, where  $\bar{\e}$ denotes the inverse of an energy scale. Using the fact that the $T \bar{T}$ operator is factorizable, one can determine many physical properties of the deformed CFT. In a coordinate system described by $x^a$ with $a=0,1$, the expectation value of $T \bar{T}$ is rewritten as a factorized form
\be
\bra T \bar{T} \ket = \fr{1}{8} \ls \fr{}{} \bra T^{ab} \ket \bra  T_{ab}  \ket - \bra T^a_a \ket^2 \rs  .
\ee
Using this factorized form, the RG equation of the deformed CFT is given by
\be		\la{Identity:scale}
\bra T^a_a \ket = - \fr{c}{24 \pi} {\cal R} - \fr{\m}{4} \ls \fr{}{} \bra T^{ab} \ket \bra T_{ab} \ket - \bra T^a_a \ket^2 \rs ,
\ee 
where $c$ indicates the central charge of the undeformed CFT. Here the first term of the right hand side  corresponds to the conformal anomaly of the undeformed CFT, while the last two terms indicate the contribution from the $T \bar{T}$ deformation.

If we represent the parameters of the deformed CFT in terms of ones appearing in the dual gravity \cite{McGough:2016lol,Donnelly:2018bef}
\be			\la{relation:parameters}
c = \fr{3 R}{2 G}  \quad {\rm and} \quad \m = 16 \pi G R ,
\ee
interestingly, the RG equation is perfectly matched to the Hamiltonian constraint of the dual gravity 
\be
0 &=& \fr{8 \pi G}{\sqrt{-g}} \lb \pi^{ab} \pi_{ab} - \ls \pi^a_a \rs^2 \rb + \fr{\sqrt{-g}}{8 \pi G} \ls {\cal R} + \fr{2}{R} \rs  .
\ee
On the dual gravity side, $g_{ab}$ denotes an induced boundary metric of a three-dimensional AdS space. In this case, $\pi^{ab} = \sqrt{-g} \ls T^{ab} - \fr{2}{\m} g^{ab} \rs$ is a conjugate momentum of the boundary metric. 

Assuming that the deformed CFT lives on an Euclidean two-dimensional sphere with a metric $ds^2 = r^2 (d \th^2 + \sin^2 \th d\ph^2)$, its partition function can be evaluated by solving the above RG equation. Noting that the energy-momentum tensor has the form of $T_{ab} = \a g_{ab}$, the RG equation determines the unknown $\a$ to be  \cite{Donnelly:2018bef}
\be
\a = \fr{2}{\m} \ls 1 - \sqrt{1 + \fr{c \m}{24 \pi r^2}} \rs .
\ee
This solution together with an appropriate boundary condition, $\log Z =0$ at $r=0$, fixes the partition function as the following form
\be
\log Z = \fr{c}{3} \sinh^{-1} \ls \sqrt \fr{24 \pi}{c \m} r \rs  + \fr{8 \pi}{\m} \ls r \sqrt{\fr{c \m}{24 \pi} + r^2} - r^2 \rs.
\ee
Moreover, the obtained partition function determines the entanglement entropy of the deformed CFT on ${\bf S}^2$
\be			\la{result:CFTentanglemententrop}
S_E =  \ls 1 - \fr{r}{2} \fr{d}{dr} \rs \log Z = \fr{c}{3} \sinh^{-1} \ls \sqrt \fr{24 \pi}{c \m} r \rs ,
\ee
where the two antipodal points of the two-dimensional sphere were taken as the entangling surface. In Ref. \cite{Donnelly:2018bef}, it has been shown that the entanglement entropy in \eq{result:CFTentanglemententrop} can be reproduced by the holographic calculation in a global AdS space with a finite UV cutoff. In this case, however, the role of the finite UV cutoff is not clear. 

In order to understand the effect of the finite UV cutoff more concretely, let us expand the entanglement entropy in the large $r$ limit ($r \gg \sqrt{c \m}$), we obtain
\be		\la{result:Donnelly}
S_E = \fr{R}{2 G} \log \fr{2 r}{R}  +  \fr{R^3}{8 G r^2 }   + \cdots ,
\ee
where the second term indicates a subleading correction caused by the finite UV cutoff denoted by $r$, which is associated with the $T \bar{T}$ deformation discussed before. Rewriting the entanglement entropy in terms of the field theory parameters in \eq{relation:parameters}, the resulting form becomes
\be		\la{result:EEonS2}
S_E = \fr{c}{3} \log \ls  \sqrt \fr{96 \pi}{c \m} r \rs + \fr{c^2 \m}{288 \pi r^2} ,
\ee
where the higher order corrections of $\m$ are ignored. For the comparison with the well-known entanglement entropy form of a two-dimensional CFT \cite{Ryu:2006bv,Ryu:2006ef}, we relate  the radius of ${\bf S}^2$ to the size of the subsystem and identify the dimensionful coupling with a shifted UV cutoff as follows:
\be			\la{result:identification}
\bar{l}=\pi r  \quad {\rm and} \quad \bar{\e}= \sqrt{c \m \pi /96 } .
\ee
We will show later that these identifications are correct by calculating the holographic entanglement entropy in a Poincare cutoff AdS$_3$ space rather than a global one used in Ref. \cite{Donnelly:2018bef}. For $\bar{l} \gg \bar{\e}$, then, the resulting entanglement entropy of the deformed CFT is expanded into
\be		\la{HEE:deformed}
S_E = \fr{c}{3} \log    \fr{\bar{l}}{\bar{\e}}   + \fr{ c }{3} \fr{\bar{\e}^2}{\bar{l}^2} + {\cal O} \ls \bar{\e}^4 \rs .
\ee
Here the first term is exactly the entanglement entropy of the undeformed CFT measured at the shifted UV energy scale ($\sim 1/\bar{\e}$) and the second term indicates the effect of the shifted UV cutoff which is associated with the previous $T \bar{T}$ deformation. This is usually an expected form for the two-dimensional deformed CFT. When the UV cutoff approaches zero, the entanglement entropy of the undeformed CFT is reproduced. Intriguingly, the resulting entanglement entropy of the $T \bar{T}$ deformation shows the almost same form as the one we obtained by the boost transformation.

It was well known that the holographic entanglement entropy evaluated in the Poincare AdS$_{d+1}$ space is also associated with that of the dual field theory defined on ${\bf S}^d$ due to the conformal symmetry \cite{Casini:2011kv,Pufu:2016zxm,Casini:2015woa}. Now, let us discuss how we can rederive the  $T \bar{T}$-deformed entanglement entropy from a Poincare cutoff AdS$_{3}$ space. In the Poincare cutoff AdS space where the position of the cutoff is denoted by $\bar{\e}$, the holographic entanglement entropy is again governed by
\be
S_E = \fr{R}{4 G} \int_{- \bar{l}/2}^{\bar{l}/2} dx \ \fr{\sqrt{1 + \bar{z}'^2}}{\bar{z}}  ,
\ee
where the subsystem size $\bar{l}$ is measured at the energy scale $\sim 1 / \bar{\e}$. Even in this case, the turning point of the minimal surface can be easily determined by the subsystem size, $l$, measured in the limit of $\bar{\e} \to 0$. Noting that the geodesic of the minimal surface follows a semicircle in \eq{solution:rescaled}, $\bar{l}$ is associated with $l$ via \eq{relation:oldnewsubsystme}, which in the previous section has been used to find the relation between the subsystem sizes of boosted and unboosted systems. As a result, the holographic entanglement entropy measured at the energy scale $\sim 1 / \bar{\e}$ results in
\be			\la{result:holoentanglementent}
S_E =  \fr{c}{3}   \log \lb \fr{\bar{l}}{2 \bar{\e}}  \left( 1 + \sqrt{1+\frac{4 \bar{\epsilon }^2}{\bar{l}^2}}   \rs \rb   .
\ee
Intriguingly, this holographic entanglement entropy obtained in a Poincare cutoff AdS$_3$  is perfectly matched to the field theory result \eq{result:CFTentanglemententrop} under the previous identifications in \eq{result:identification}. Applying the definition of the $c$-function studied in the previous boost transformation, the $c$-function of the  $T \bar{T}$ deformation and its RG flow are described by
\be
\bar{c} (\bar{\e}) &=& \fr{c \ \bar{l}}{\sqrt{\bar{l}^2 + 4  \bar{\e}^2} } ,   \\
\fr{\pa \bar{c}}{\pa \log \bar{\e}}  &= & -  \fr{ 8 c \ \bar{l} \ \bar{\e}^2}{ \ls \bar{l}^2 + 4  \bar{\e}^2 \rs^{3/2} }  < 0.  
\ee
The boost and $T \bar{T}$ deformations studied here, intriguingly, show the very similar behavior except that the $c$-function of the theory deformed by the boost transformation decreases more rapidly (see Fig. 1).


\section{Discussion}

In this paper, we have investigated the effect of the boost transformation and $T \bar{T}$ deformation on the entanglement entropy. In general, the boost transformation is trivially acting on the underlying CFT and QFT. Even in this case, we showed that the boost transformation gives rise to a nontrivial effect on the entanglement entropy. The reason is that the boost transformation modifies the size and shape of the entangling surface due to the length contraction caused. In general, the boost and scale transformations are not well-defined transformations in a finite subsystem introduced to define the entanglement entropy. In this work, we explicitly showed that the entanglement entropy of a boosted system can be reinterpreted as the shift of the UV cutoff for a two-dimensional QFT, similar to the $T \bar{T}$ deformation.

According to the AdS/CFT correspondence, the dual gravity of an undeformed CFT is described by a one-dimensional higher AdS geometry. However, since boosting the entanglement entropy of a two-dimensional CFT causes the shift of the UV cutoff, the dual geometry of the boosted two-dimensional CFT becomes a cutoff AdS space with the boundary at a finite distance. It has already been known that the $T \bar{T}$ deformation also shows the similar behavior. After calculating the holographic entanglement entropy, which is valid in the entire range of the energy scale, we investigate the entanglement entropy and $c$-function relying on the energy scale denoted by the shifted UV cutoff. Thanks to the reinterpretation of the boost as the shift of the UV cutoff, we can investigate the change of the $c$-function in the entire range of the RG scale. The boost transformation of the entanglement entropy shows that a two-dimensional CFT at a UV fixed point flows to another CFT at an IR fixed point, as expected by Zamolodchikov's $c$-theorem. In this case, the UV fixed point is stationary and the $c$-function monotonically decreases along the RG flow.

We also investigated whether the similar reinterpretation of the shift of the UV cutoff is possible even for higher dimensional theories and found that such a reinterpretation is only possible for a two-dimensional CFT and its dual gravity. As a consequence, reinterpreting the boost as the shift of the UV cutoff may be a specific feature of a two-dimensional theory. In spite of this fact, the boost transformation can provide an important result about the entanglement entropy. The length contraction caused by the boost usually modifies the size and shape of the entangling surface. For example, a disk-shaped region in the unboosted system can be changed into an ellipsoidal one. We showed that the area law of the entanglement entropy is still satisfied even after the boost transformation.

Applying the method used in the boosted system, we also studied the entanglement entropy of a CFT deformed by a $T \bar{T}$ operator. We showed that the $T \bar{T}$ deformation also leads to the similar feature to the boost. In addition, we holographically rederived the same entanglement entropy deformed by the $T \bar{T}$ operator from the cutoff AdS space in the Poincare patch. This method can be easily generalized to a higher dimensional case. We hope to report more results in future work.

\vspace{1cm}

{\bf Acknowledgement}

C. Park was supported by Basic Science Research Program through the National Research Foundation of Korea funded by the Ministry of Education (NRF-2016R1D1A1B03932371).

\vspace{1cm}


\end{document}